# Influence of X-ray Irradiation on the Magnetic and Structural Properties of Gadolinium Silicide Nanoparticles for Self-Regulating Hyperthermia

**Samantha E. Smith[1], Santiago Bermudez[2], Pavan Chaitanya[2], Zoe Boekelheide[3], Jessika Rojas Marin[2] and Ravi L. Hadimani[2, 4, 5*]**

[1]Department of Chemical and Life Sciences Engineering, Virginia Commonwealth University, Richmond, VA 23284

[2]Department of Mechanical and Nuclear Engineering, Virginia Commonwealth University, Richmond, VA 23284

[3]Department of Physics, Lafayette College, Easton, PA 18042

[4]Department of Biomedical Engineering, Virginia Commonwealth University, Richmond, VA 23284

[5]Department of Electrical and Computer Engineering, Iowa State University, Ames, IA 50011

*Corresponding author's E-mail: rhadimani@vcu.edu

**Abstract:** Magnetic hyperthermia treatment (MHT) utilizes heat generated from magnetic nanoparticles (MNPs) under an alternating magnetic field (AMF) for therapeutic applications. Gadolinium silicide ($Gd_5Si_4$) has emerged as a promising MHT candidate due to its self-regulating heating properties and potential biocompatibility. However, the impact of high-dose X-ray irradiation on its magnetic behavior remains uncertain. This study examines $Gd_5Si_4$ nanoparticles exposed to 36 and 72 kGy X-ray irradiation at a high-dose rate (120 Gy/min). While X-ray diffraction, scanning electron microscopy, and energy dispersive spectroscopy confirm no structural or compositional changes, transmission electron microscopy reveals localized lattice distortions, along with observable changes in magnetic properties, as evidenced in magnetization vs. temperature and hysteresis measurements. Despite this, magnetocaloric properties and specific loss power (SLP) remain unaffected. Our findings confirm the stability of $Gd_5Si_4$ under high-dose X-ray irradiation, supporting its potential for radiotherapy (RT) and magnetocaloric cooling in deep-space applications.

## 1. Introduction

Magnetic hyperthermia (MH) has emerged as a promising cancer treatment in nanomedicine, drawing significant research interest in this rapidly evolving field [1]. Originally proposed in 1957 as an effective strategy to detect and eliminate metastatic cells in lymph nodes that may be missed during surgical cancer removal procedures [2], MH is now available in Germany as a treatment option for glioblastoma and prostate cancer [3], [4], [5]. MHT involves delivering magnetic nanoparticles (MNPs) into the body and guiding them to a cancerous tumor [6]. Once the magnetic nanoparticles accumulate in the cancerous tumor, an alternating magnetic field (AMF) is applied at frequencies restricted to the therapeutic range of hundreds of kHz. This rapid switching of the magnetic particle moments results in energy dissipation as heat, equal to the area enclosed in the magnetic hysteresis loop (M(H)) during each field cycle [4], raising the local temperature to ~42–46 °C [7]. This temperature increase, due to the heat produced, damages cancerous cells without harming non-cancerous cells. Moreover, this effect can potentially be used to enhance other therapies, such as radiation and chemotherapy [8].

MNPs such as iron oxide are currently being used due to their high specific loss power (SLP) and their approval under the US Food and Drug Administration [9], [10], [11]. Additionally, other platforms, such as $Gd_5Si_4$, are being explored for their potential in magnetic hyperthermia (MH) applications. The high magnetization and stability of $Gd_5Si_4$, combined with its promising biocompatibility, make it an exciting prospect for magnetic hyperthermia treatment [4]. Furthermore, its potential use as a cancer treatment in combination with radiation therapy presents a range of parameters to consider when evaluating MNPs. Proposed as a promising candidate for magnetic hyperthermia treatment, $Gd_5Si_4$ continues to be investigated for its ability to adjust its transition temperature to meet the criteria for self-controlled hyperthermia [12], [13], as well as being an excellent T2 contrast agent due to its ferromagnetic phase at healthy human homeostatic temperature [14].

Magnetic materials, especially MNPs, have shown to degrade when exposed to high-energy X-ray radiation [15], [16]. Given the potential for $Gd_5Si_4$ nanoparticles to experience substantial radiation exposure in clinical settings or in high radiation environments, investigating the effects of ionizing radiation on their intrinsic magnetic properties is crucial. This exploration will help determine the viability of $Gd_5Si_4$ as a candidate for coupling MHT with RT [17].

In this study, we investigate the effects of X-ray radiation on the magnetic and structural properties of $Gd_5Si_4$ using various characterization techniques. X-ray diffraction (XRD) and transmission electron microscopy (TEM) were employed to analyze the morphology and crystal structure of the material before and after irradiation. The elemental composition was assessed using energy-dispersive X-ray spectroscopy (EDS) in conjunction with scanning electron microscopy (SEM). Magnetization (M) as a function of temperature (T), along with hysteresis behavior, was measured using a vibrating sample magnetometer (VSM) to determine transition temperatures and magnetization values. Additionally, specific loss power (SLP) was measured to evaluate magnetic MHT performance. These characterization methods provide a comprehensive assessment of $Gd_5Si_4$ as a magnetic nanoparticle for MHT following radiation exposure.

## 2. Experimental procedure

***2.1. Material Synthesis.*** $Gd_5Si_4$ used in this study was manufactured using a polycrystalline sample synthesized via arc melting of a stoichiometric mixture of Gd (99.9% purity) and Si (Cerac Inc., USA, >99.999% purity). The samples were melted under an Ar atmosphere and remelted six times to ensure homogeneity [18]. Commercial-grade Gd was used to minimize impurities in the $Gd_5Si_4$ matrix, such as $Gd_5Si_3$ and GdSi [19], [20].

To achieve sub-micrometer particles of $Gd_5Si_4$, the as-cast material was initially ground in an agate mortar and then sieved to collect powders with a particle size of 50 μm or smaller. Further particle size reduction was achieved through high-energy ball milling using a SPEX 8000M mill. All milling and material manipulations were carried out in a glove box under an argon atmosphere. Milling was performed with a ball-to-powder weight ratio of ~5:1, using two balls with an 11 mm diameter and four balls with a 4 mm diameter. Poly (ethylene glycol) (PEG, molecular weight: 8000 Da) was added at 10% by weight as a surface modifier during milling.

A two-step milling process was employed. First, the $Gd_5Si_4$ and PEG mixture was milled for 1 hour. After this step, 5 mL of heptane was added to prevent particle binding due to localized melting and to retain crystallinity [21]. The second milling step involved an additional 1 hour of milling, during which the heptane evaporated, yielding a fine $Gd_5Si_4$ powder mixed with PEG [22].

***2.2. X-ray Irradiation.*** X-rays were produced using a compact cabinet irradiator (X-RAD 225XL, Precision X-Ray Inc.) equipped with a tungsten target. The characteristic X-ray emissions were 59.32 keV (Kα1), 57.98 keV (Kα2), 67.24 keV (Kβ1), 69.10 keV (Kβ2), and 66.95 keV (Kβ3). The system operated at an acceleration voltage of 225 kV and a current of 13.33 mA. No filters or collimators were applied. The irradiation experiments were conducted at a dose rate of 120 Gy/min for 5 and 10 hours, achieving absorbed doses of approximately 36 kGy and 72 kGy, respectively. X-rays were emitted in a cone beam at an angle of 40 degrees, as shown in **Figure 1**.

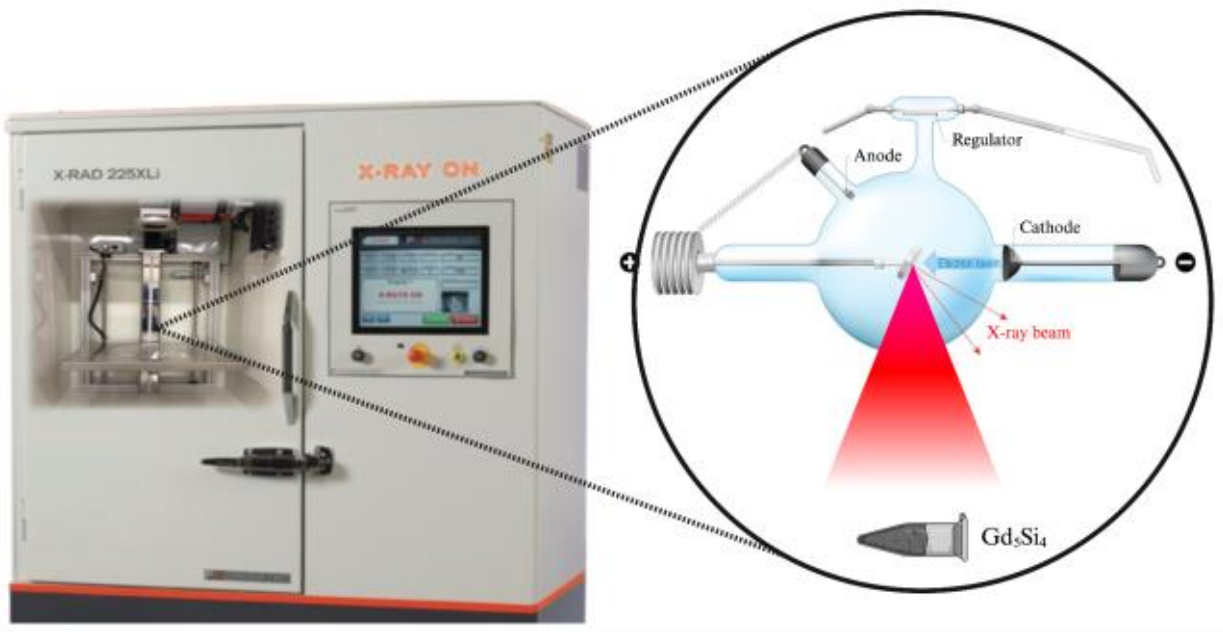

**Figure 1.** *Experimental setup for the irradiation of $Gd_5Si_4$ nanoparticles.*

***2.3. X-ray Diffraction (XRD).*** $Gd_5Si_4$ nanoparticles were analyzed for morphology and crystal structure using X-ray diffraction (XRD, Panalytical MPD X'Pert Pro). The device uses a HyPix-400 MF 2D detector (receiving slit size: 13 mm) and a 40 kV, 15 mA Cu X-ray source (divergence slit: 1.25°, scatter slit size: 8 mm).

***2.4. Scanning Electron Microscopy (SEM).*** SEM and EDS were performed using a Hitachi SU70 Field Emission Scanning Electron Microscope (FESEM) with a field emission gun (FEG) for high-resolution imaging. An accelerating voltage of 15 kV was selected to optimize the detection of expected elements, particularly heavier elements that may require higher voltages for ionization. A working distance of 10 mm was maintained for optimal focus and resolution. For select areas, elemental mapping was performed to visualize the spatial distribution of elements across the sample surface.

***2.5. Transmission Electron Microscopy (TEM).*** The crystalline structure and particle size were characterized using a JEOL JEM-F200 transmission electron microscope (TEM) operating at an accelerating voltage of 200 kV. For sample preparation, a 1 mg/mL solution of nanoparticles was diluted in DI water at a 1:20 ratio, followed by redispersion in a sonication bath to prevent overheating. Afterward, 10 μL of the diluted nanoparticle suspension was deposited onto a 300-mesh copper grid and allowed to dry overnight.

***2.6. X-ray Photoelectron Spectroscopy (XPS).*** X-ray photoelectron spectroscopy (XPS) measurements were conducted using a PHI VersaProbe III Scanning XPS Microprobe equipped with a monochromatic, micro-focused Al Kα (1486 eV) X-ray source. All measurements were performed under ultrahigh vacuum conditions ($<10^{-8}$ Pa). To compensate for charge buildup, an electron flood gun was utilized. Binding energy calibration was achieved using the C1s peak of adventitious carbon at 284.8 eV.

***2.7. Magnetic Characterization.*** A vibrating sample magnetometer (VSM, Quantum Design, 3 T VersaLab) was used to assess the magnetic properties of the particles. The magnetization (M) of $Gd_5Si_4$ samples was measured over a temperature range of 200–400 K. Measurements were performed under an applied magnetic field ranging from -3 T to 3 T.

***2.8. Magnetic Hyperthermia Measurements.*** Measurements of the specific loss power (SLP) were conducted on particles dispersed in 500 μL of $H_2O$ in a flat-bottomed glass vial at a concentration of approximately 25 mg/mL at room temperature. An Ambrell EasyHeat AMF generator with an eight-turn, water-cooled solenoidal coil (**Figure 2**)was operated at a frequency of 226 ± 1 kHz. The field strength, $\mu_0H_{max}$, at the center of the coil was 0.035 T, calibrated using a Fluxtrol alternating magnetic field probe. The applied field strength $\mu_0H_{max}$ is expressed in terms of the root mean square (RMS) field strength: $\mu_0H_{max} = \sqrt{2}\,\mu_0H_{max}$ since the field amplitude varies slightly with each cycle. The sample was placed at the center of the solenoid, and the AMF was applied. The temperature as a function of time, T(t), was monitored using a 36-gauge E-type thermocouple. E-type thermocouples can be used in AMF environments if they are sufficiently thin (36 gauge), as this minimizes eddy current heating [23].

The AMF was applied until a temperature change of at least 0.5 °C was achieved so that the temperature increase could be modelled by a linear fit. Then, SLP was calculated from the initial slope of the temperature rise upon application of the AMF, according to the equation:

$$SLP = \frac{cm_w}{m_p}\left(\frac{dT}{dt}\right) \qquad (1)$$

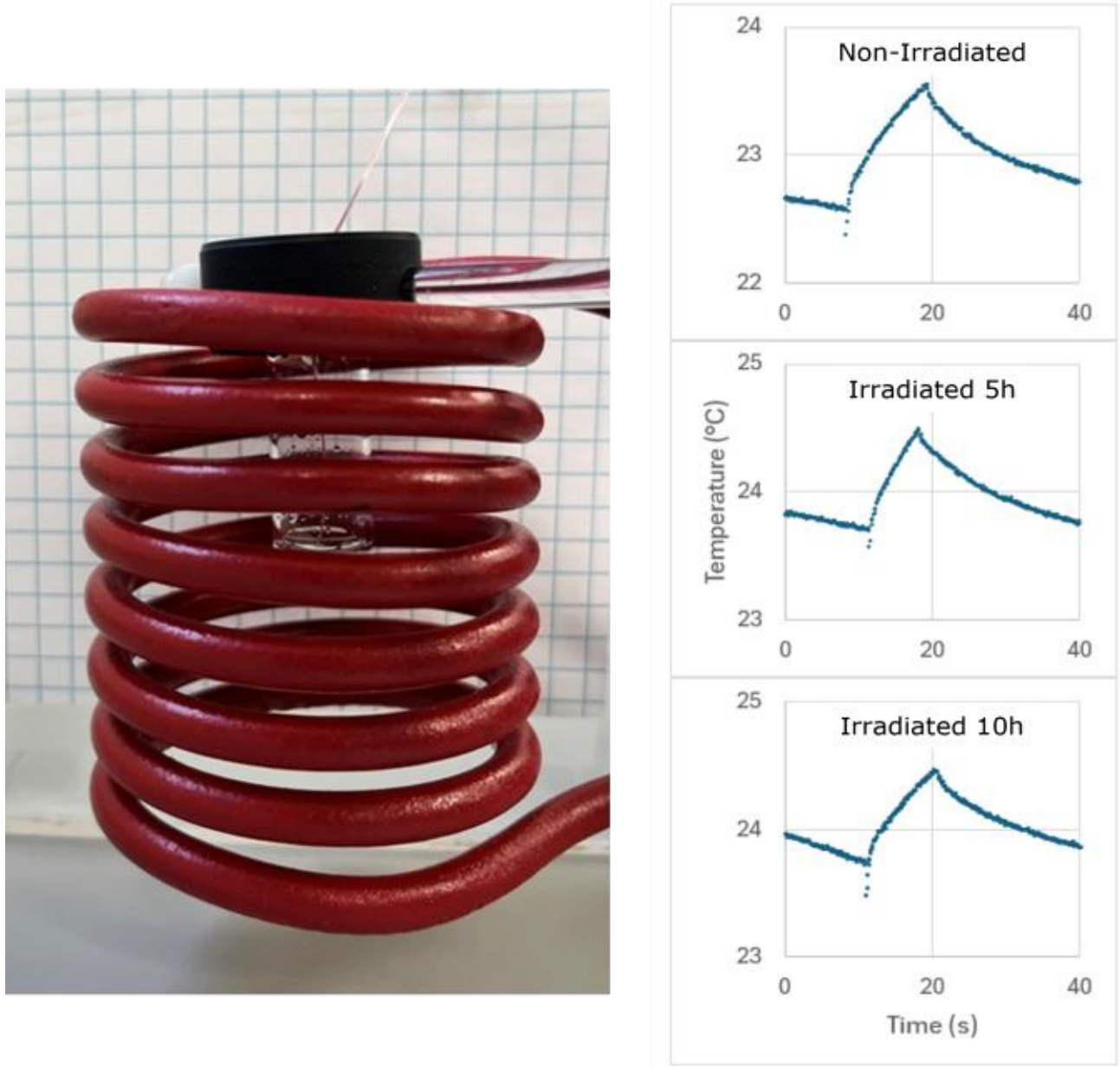

**Figure 2.** *Left: Solenoidal AMF coil with vial and 36-gauge E-type thermocouple. Right: Example heating curve from $Gd_5Si_4$ nanoparticle sample.*

where *c* is the specific heat capacity of water, $m_w$ is the mass of water, and $m_p$ is the mass of the sample (particles). It is assumed that the heat capacity of the sample is dominated by the water. The same measurement was performed on 500 μL of water to determine the background *dT/dt*, which was significantly smaller than the sample signal. This background *dT/dt* was subtracted from the sample *dT/dt* to obtain the slope specific to the magnetic samples. Losses within this temperature range were also subtracted to obtain a corrected slope [24]. The uncertainty in the SLP measurement is 15%, as determined through multiple trials.

## 3. Results and discussion

### 3.1. XRD Analysis

**Figure 3** shows the XRD spectra of $Gd_5Si_4$ nanoparticles before and after 5 and 10 hours of irradiation. The results indicate that $Gd_5Si_4$ (ICSD: 84083) nanoparticles predominantly crystallize in the orthorhombic structure, with space group Pnma (No. 62), and exhibit minor impurity peaks corresponding to GdSi (ICSD: 154515) and $Gd_5Si_3$ (ICSD: 636425). These impurities are known to be challenging to eliminate, even after a prolonged heat treatment. Since GdSi is antiferromagnetic and $Gd_5Si_3$ is paramagnetic at room temperature with nearly zero spontaneous magnetization, their presence in small quantities is unlikely to have a significant impact on the magnetic properties of $Gd_5Si_4$ [25], [26], [27].

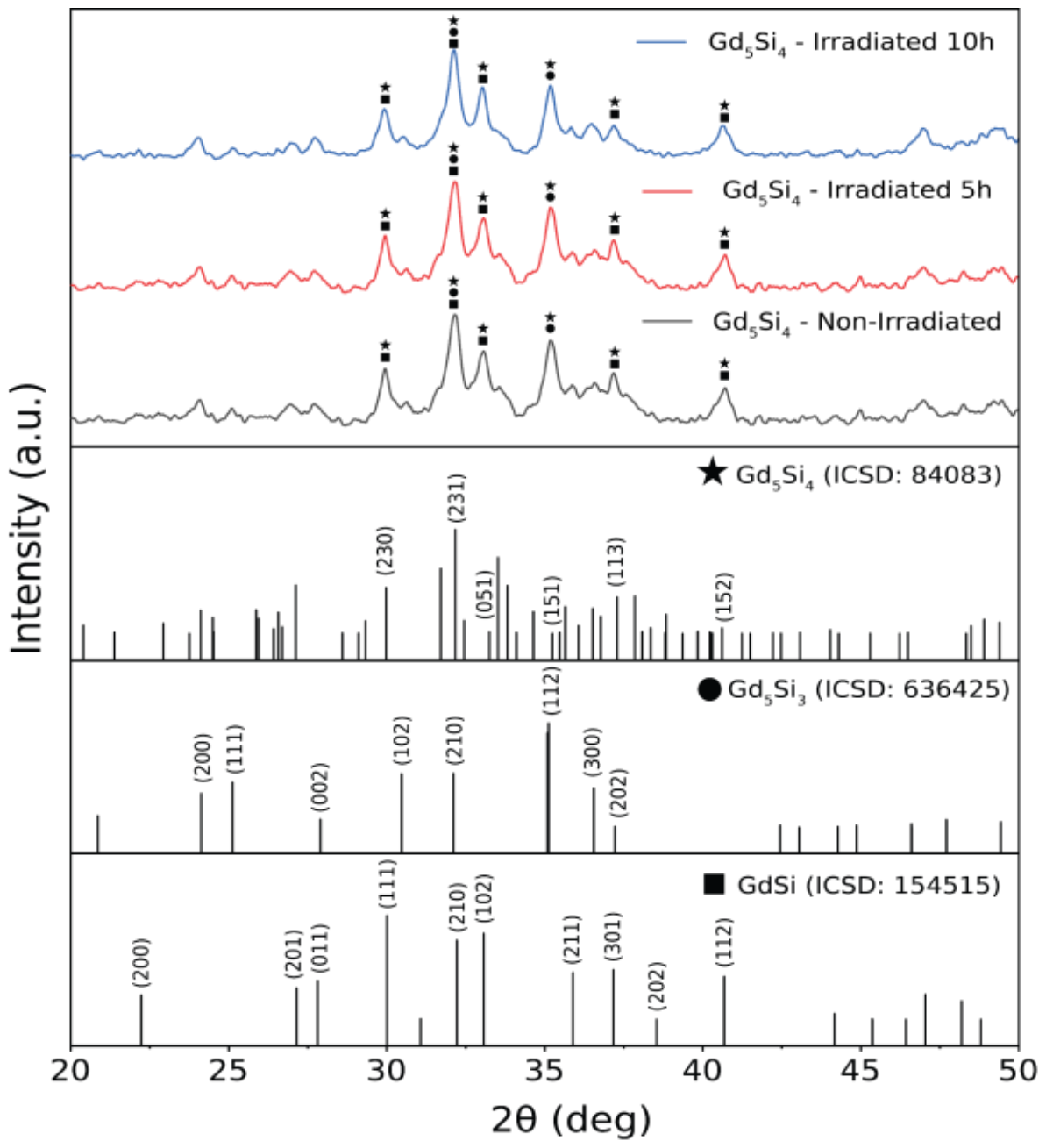

**Figure 3.** *Room-temperature XRD Result Patterns for $Gd_5Si_4$ before and after X-ray irradiation treatment (225 kV, 13.33 mA) for 5 and 10 h.*

The XRD spectra before and after irradiation did not show substantial differences in peak positions; however, a slight broadening of the most intense diffraction peaks was observed after X-ray irradiation. This broadening can be attributed to the introduction of defects, which disrupt the periodic arrangement of atoms. These defects generate microstrain within the crystal lattice, causing minor variations in atomic spacing. Since these defects typically occur in low concentrations, they do not significantly alter the overall crystal structure and produce only minor changes in the diffraction pattern, making them difficult to detect. Some defects are already introduced by the ball-milling process, and the number of defects is slightly increased by irradiation.

## 3.2. SEM/EDX

The morphological features of the $Gd_5Si_4$ nanoparticles, as revealed by scanning electron microscopy (SEM), showed no significant variations between the non-irradiated and irradiated samples (**Figure 4a–c**). In both irradiated samples, the particles maintained their morphology, indicating high resistance to surface degradation or structural reorganization under X-ray exposure. At lower magnification **(Figure 4a)**, the sample appears as an agglomeration of irregularly shaped particles with varying sizes, indicating heterogeneous grain morphology. As magnification increases **(Figures 4b and 4c)**, the particle boundaries become more distinct, and larger grains with smoother surfaces are evident, suggesting a degree of sintering or grain growth.

Elemental analysis conducted via Energy Dispersive X-ray Spectroscopy (EDS) further confirmed the compositional stability of the material (**Figure 4d–f**). From **Table 1,** the elemental distribution, expressed in weight percent (wt.%), remained consistent with only minor fluctuations observed over time. The non-irradiated sample exhibited a composition of 0.90 wt.% oxygen (O), 12.82 wt.% silicon (Si), and 86.28 wt.% gadolinium (Gd). After 5 hours of X-ray exposure, oxygen decreased slightly to 0.63 wt.%, silicon to 11.79 wt.%, and gadolinium increased modestly to 87.58 wt.%. Following 10 hours of irradiation, the composition remained comparably similar at 0.94 wt.% O, 11.66 wt.% Si, and 87.40 wt.% Gd.

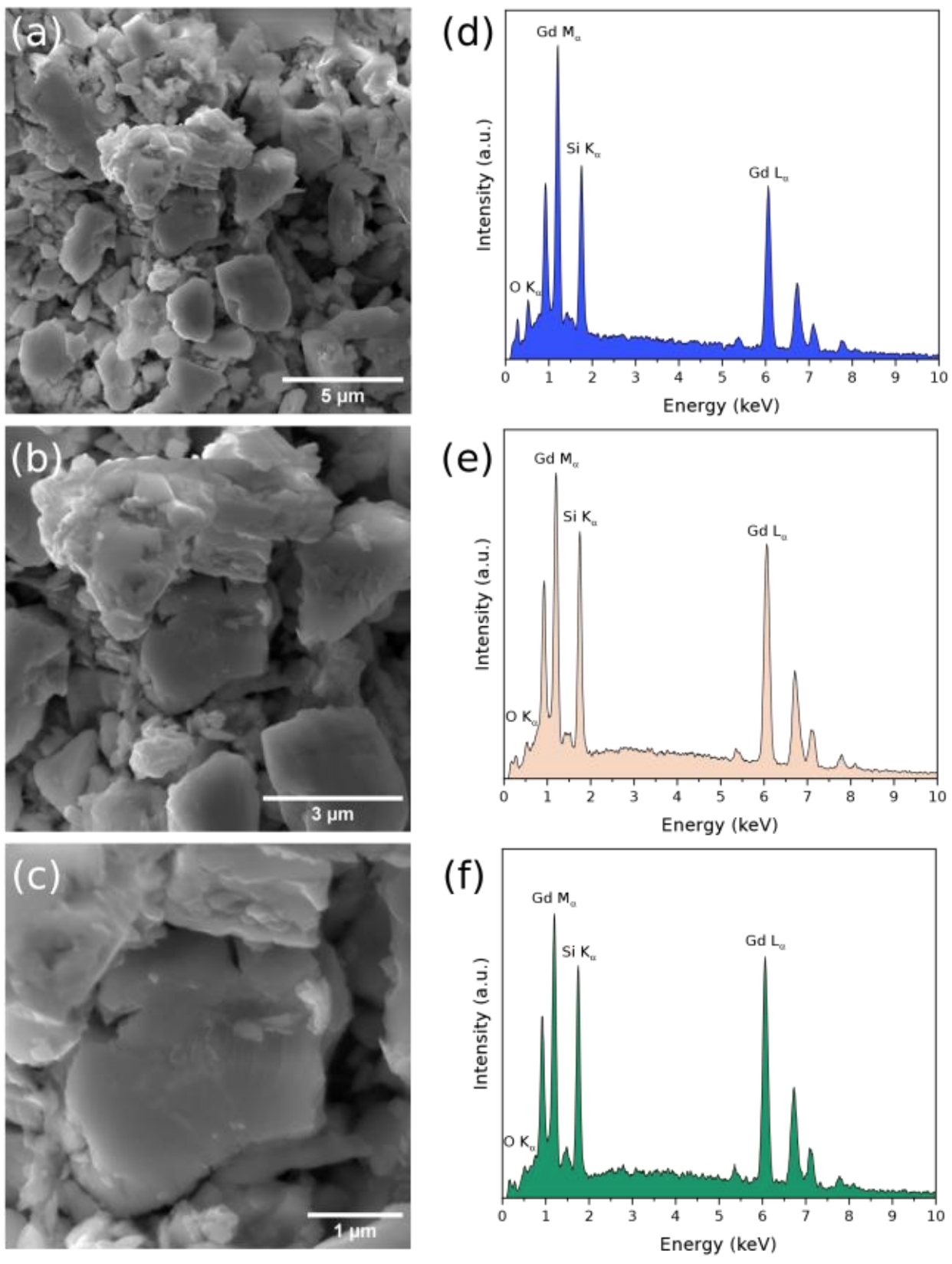

**Figure 4.** *SEM micrographs of $Gd_5Si_4$ at increasing magnifications: (a) low, (b) medium, and (c) high. Corresponding EDX spectra for (d) non-irradiated, (e) irradiated for 5 hours, and (f) irradiated for 10 hours.*

These minor variations fall within typical experimental uncertainty and are not indicative of significant compositional changes, which confirms that $Gd_5Si_4$ nanoparticles exhibit excellent chemical and structural stability under extended X-ray irradiation. For further reference, the elemental distribution is also presented in atomic percent (at.%) in **Table 2**.

**Table 1.** *Elemental composition in wt.% of non-irradiated and irradiated samples at different time intervals.*

| Element | Composition in wt.% | | |
|---|---|---|---|
| | Non-Irradiated | Irradiated 5h | Irradiated 10h |
| **O** | 0.90 | 0.63 | 0.94 |
| **Si** | 12.82 | 11.79 | 11.66 |
| **Gd** | 86.28 | 87.58 | 87.40 |

**Table 2.** *Elemental composition in at.% of non-irradiated and irradiated samples at different time intervals.*

| Element | Composition in at.% | | |
|---|---|---|---|
| | Non-Irradiated | Irradiated 5h | Irradiated 10h |
| **O** | 5.30 | 3.87 | 5.69 |
| **Si** | 43.00 | 41.31 | 40.33 |
| **Gd** | 51.70 | 54.82 | 53.99 |

Even though EDS is a semi-quantitative technique, meaning that while it provides valuable insights into elemental composition, slight variations may arise due to factors such as sample surface roughness and detector efficiency, the minimal variation in composition over extended irradiation highlights the material's structural and compositional resilience, making it suitable for medical applications. The minor variations in peak intensity may reflect localized compositional differences or differences in grain orientation and size. These results indicate that $Gd_5Si_4$ nanoparticles has potential for use in multi-modal cancer treatments.

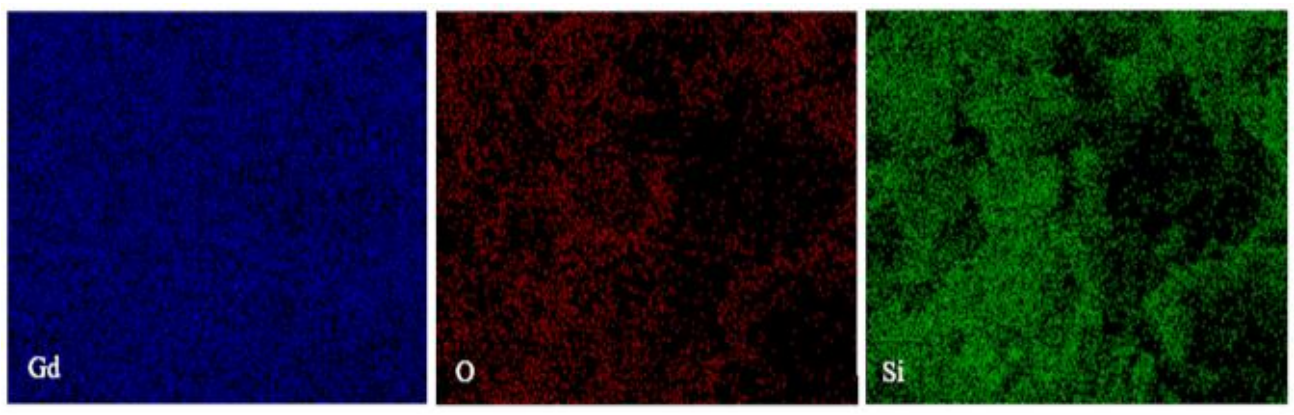

**Figure 5.** EDS abundance maps for the selected elements

**Figure 5** displays the elemental distribution maps obtained via EDS for the synthesized sample, highlighting the spatial presence of gadolinium (Gd), oxygen (O), and silicon (Si). The elemental maps indicate a relatively homogeneous distribution of all three elements across the analyzed area, confirming the successful incorporation of the constituent elements. The Gd map (blue) shows uniform dispersion throughout the matrix, suggesting consistent integration of the rare-earth element in the host structure. Similarly, the Si map (green) indicates a widespread presence, supporting the formation of a silicide or oxide-silicate phase. The oxygen map (red), while slightly less dense in some regions, still shows good coverage, implying partial surface oxidation or the formation of an oxide matrix around Gd and Si components.

In particular, the high gadolinium content and stability make gadolinium neutron capture therapy (GdNCT) a promising approach for cancer treatment modalities, where consistent performance under radiation is critical. This stability ensures the material retains its therapeutic efficacy and structural integrity, even under prolonged exposure to high-energy radiation. Although neutron irradiation is not explored in this study, $Gd_5Si_4$ could be a good candidate for neutron capture therapy, and further research should investigate the effects of neutron irradiation on this material.

## 3.3. XPS

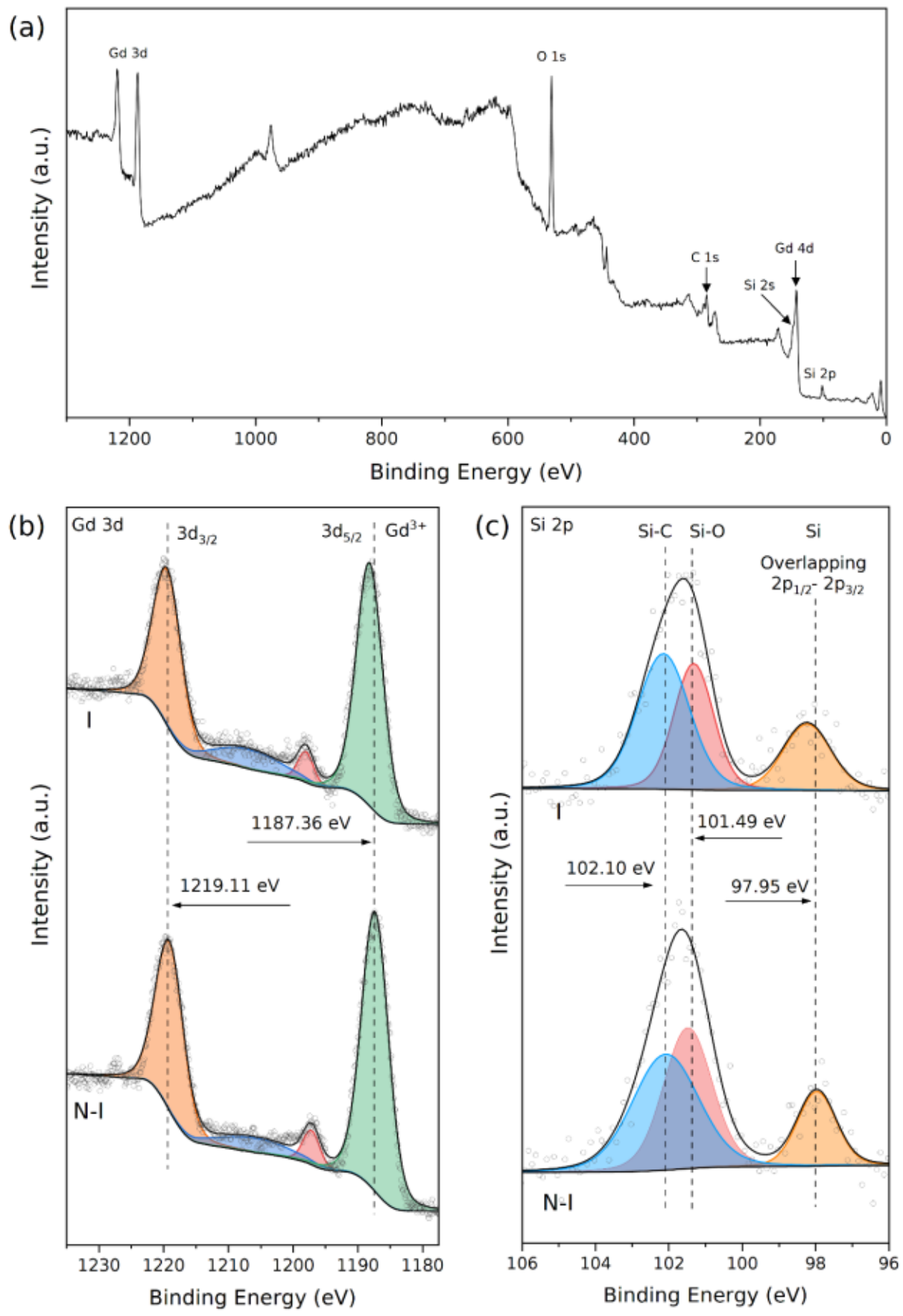

**Figure 6.** *(a) XPS survey scan of $Gd_5Si_4$. (b) High-resolution Gd 3d spectra. (c) High-resolution Si 2p spectra for irradiated (I) and non-irradiated (N-I) samples.*

**Figure 6** presents the X-ray photoelectron spectroscopy (XPS) analysis of $Gd_5Si_4$ samples, comparing non-irradiated and irradiated conditions. The survey scan (**Figure a**) confirms the presence of Gd, Si, O, and C. The prominent Gd 3d peak at ~1187 eV and Si 2p peak at ~98 eV indicate the presence of the primary $Gd_5Si_4$ phase. XPS survey analysis determined the surface composition to be 42.71 wt% oxygen, 20.07 wt% carbon, 20.98 wt% silicon, and 16.24 wt% gadolinium. However, the presence of the O 1s signal suggests surface oxidation, likely resulting from the formation of gadolinium or silicon oxides due to air exposure or the milling process. Additionally, the presence of silicon compounds may enhance biocompatibility, supporting the material's potential for use in biological systems [28].

High-resolution Gd 3d spectra (**Figure 6 b**) show two major doublets corresponding to Gd $3d_{5/2}$ and Gd $3d_{3/2}$, along with some satellites that may appear due to multiplet splitting and screening effects. In the non-irradiated sample (N-I), the Gd $3d_{5/2}$ peak appears at ~1187.36 eV, while in the irradiated sample (I), a slight shift is observed, suggesting possible minor changes in the local chemical environment. In addition, it is expected that the presence of surface $Gd_2O_3$ or $Gd(OH)_3$ can shift Gd 3d peaks and introduce satellite structures [29], [30], [31], [32].

The high-resolution Si 2p spectra exhibit spin-orbit components Si $2p_{3/2}$ and $2p_{1/2}$, which overlaps and were located at 97.95 eV. Deconvolution of the Si 2p region reveals two additional peaks at 101.49 eV and 102.10 eV, corresponding to the formation of surface Si–C and Si–O/Si–O–C species, respectively. The formation of oxycarbide phases on the film surface may have resulted from atmospheric exposure or during the milling process. The observed values are in good agreement with those previously reported in the literature [33].

## 3.4. TEM

**Figure a-c** shows the TEM images of $Gd_5Si_4$ nanoparticles, showing a clear tendency for smaller particles to aggregate, forming dense, interconnected clusters. This aggregation behavior likely results from the high surface energy, which promotes particle aggregation and further coalescence into larger structures. Additionally, the milling process can potentially contribute to a broad size distribution, reflecting the inherent limitations of achieving uniform nanoparticle dispersion without the use of surfactant-assisted milling. The estimated average particle size was found to be 12.21 nm, with a standard deviation of 5.16 nm. While this provides a general representation of the particle size distribution, the presence of some isolated larger particles was observed in the sample. These larger particles are likely residuals from the milling process, suggesting that not all particles were fully reduced to the desired nanoscale size.

Before irradiation, HR-TEM images (**Figure 7 d-f**) and their corresponding FFT patterns confirmed that the nanoparticles exhibited some defects, maintaining a well-ordered crystalline structure. The lattice fringes appeared continuous and uniform, with sharp diffraction spots in the FFT patterns, indicating high crystallinity and structural integrity.

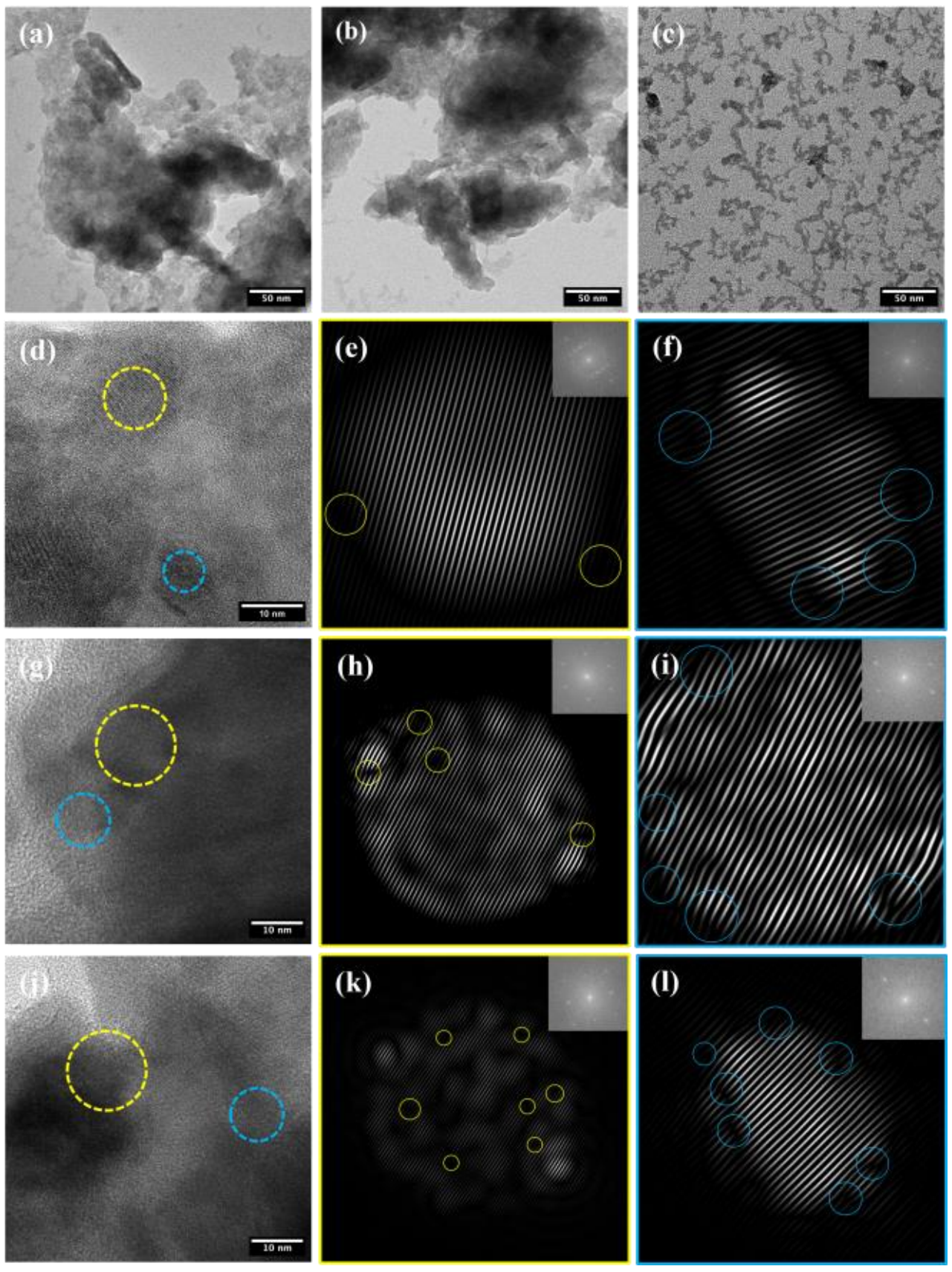

**Figure 7.** *TEM images of (a-c) $Gd_5Si_4$ nanoparticles, HR-TEM images, and corresponding FFT patterns marked in yellow and blue of $Gd_5Si_4$ nanoparticles: (d-f) non-irradiated, (g-i) after 5 h of irradiation and (j-l) after 10 h of irradiation at 225 kV and 13.33 mA.*

Upon irradiation, the nanoparticles exhibited progressive structural changes. After 5 hours of irradiation, HR-TEM images (**Figure 7 g-i**) showed localized lattice distortions and dislocations, showing the possible early stages of defect formation. While some regions retained their crystalline order, as confirmed by the corresponding FFT patterns, others displayed clear signs of structural damage. Increasing the irradiation time to 10 hours (**Figure 7 j-l**), a significant increase in defect density was observed, resulting in extended lattice damage and crystalline disorder. This was evident from the diffuse spots in the FFT patterns, indicating a breakdown of the ordered structure. The accumulation of defects introduced considerable stress and dislocations, which can disrupt the alignment of magnetic domains and potentially affect the magnetic properties by causing atomistic spin disorder [34].

While understanding the influence of particle shape and size on magnetic properties is crucial, it is even more important to investigate how the internal atomic structure within each domain contributes to these properties. The arrangement of atoms, defect distribution, and crystallographic order plays a significant role in determining magnetic behavior, affecting factors such as anisotropy, domain wall movement, and overall performance [35].

### 3.5. Magnetic Characterization

Magnetic hysteresis loops were measured at 300 K for both non-irradiated and irradiated samples (**Figure** 8 **a–b**), exhibiting ferromagnetic behavior consistent with prior literature [36], [37], [38].

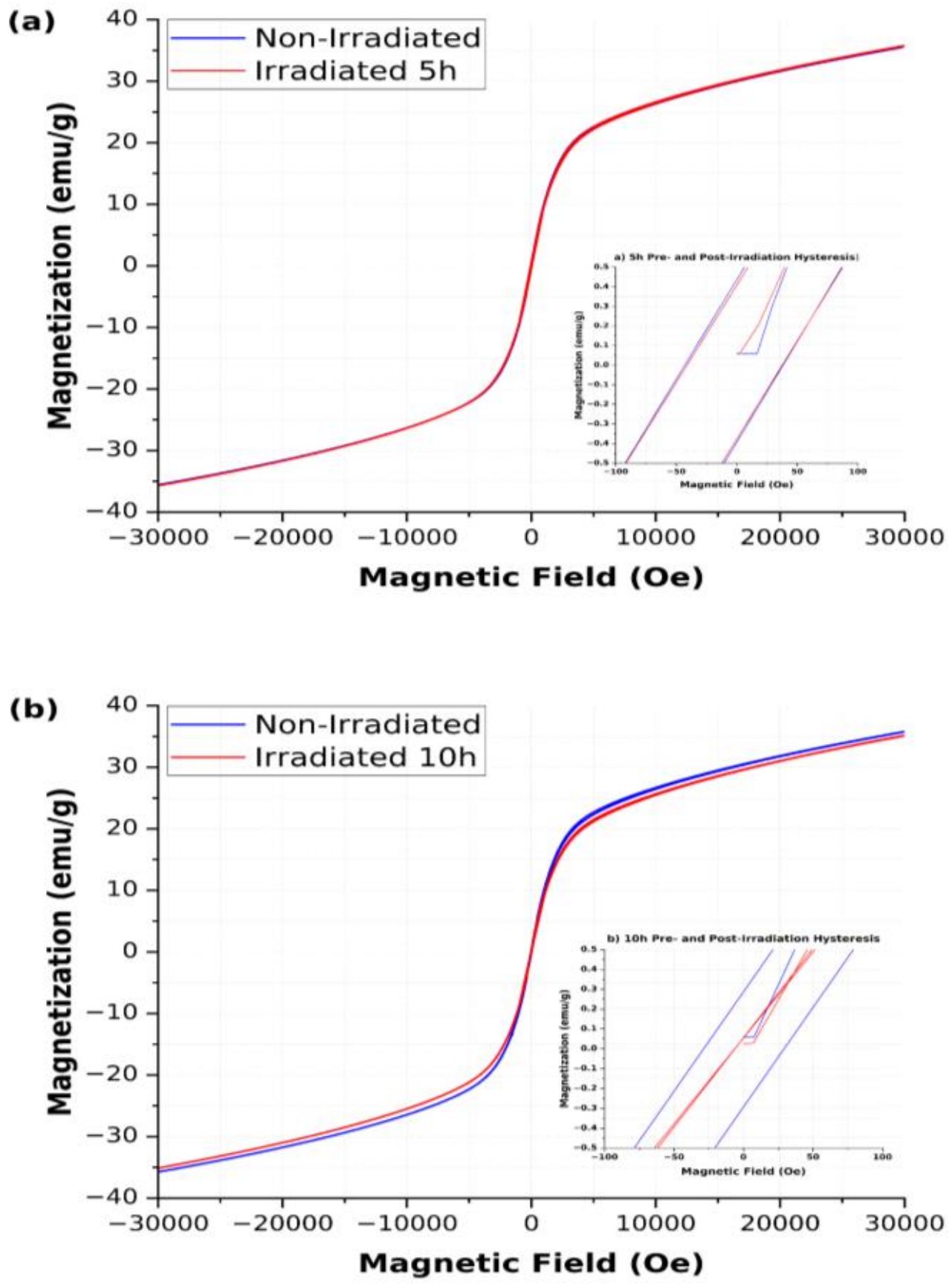

**Figure 8.** *VSM magnetic hysteresis loops at 300 K from -3 T to 3 T of non-irradiated and irradiated samples after (a) 5h and (b) 10h at 300 K.*

When the data for the 5-hour irradiated sample are overlapped (Figure 8a), the magnetization shows no significant observable change. Additionally, no significant variation in coercivity is observed, as shown in the inset of **Figure 8 a**.

A slight decrease in magnetization (M) is observed in the 10h irradiated sample (**Figure 8 b**), as seen in **Table-3** with a $M_s$ value 0.65 emu/g lower than the pre-irradiated sample. The hysteresis graphs of the non-irradiated and 10h irradiated samples (**Figure 8 b**) overlap, showcasing the observable reduction in magnetization. When examining the origin of the hysteresis graphs for **Figure 8 b**, a slight decrease in coercivity of approximately 34 Oe is observed.

**Table 3.** *Magnetic Properties of $Gd_5Si_4$ Before and After X-ray Irradiation*

| | Non-Irradiated | Irradiated 5h | Irradiated 10h |
|---|---|---|---|
| **Ms (emu/g)** | 35.72 | 35.81 | 35.07 |
| **Mr (emu/g)** | 0.43 | 0.40 | 0.05 |
| **Hc (Oe)** | 40.53 | 39.30 | 6.16 |

Changes in the magnetic properties of the $Gd_5Si_4$ nanoparticles as a function of irradiation time are summarized in **Table 3**. This data provides further insight into the material's response to prolonged X-ray exposure.

The magnetization (M) as a function of temperature (T) is presented in **Figure 9 a** for the non-irradiated sample of $Gd_5Si_4$. The pre-irradiated sample shows the material as ferromagnetic until around 334 K (60 ℃). The transition temperature peak shows a decrease from the bulk value $T_C$ of 336 K (63 ℃) [39]. This change in transition temperature showcases a decrease towards an appropriate therapeutic range for use in magnetic hyperthermia treatment of 43-45 ℃ [40], which could be further reduced by decreasing the particle size [12].

Similarly, **Figure 9 b** showcases M-T data for the sample of $Gd_5Si_4$ after exposure to 5 hours of radiation. No significant change in dM/dT between **Figure 9 a** and **Figure 9 b** is seen. Magnetization (M) as a function of temperature (T) for the 10-hour irradiated sample (**Figure 9 c**) a transition temperature of approximately 57 ℃, which remains lower than the bulk $Gd_5Si_4$ transition temperature of 336 K (63℃) [39]. When the material is exposed to radiation for 5 hours no significant transition temperature change is seen, and when exposed to 10 hours of radiation, a decrease in transition temperature of 3 K/℃ is observed. The change in the transition temperature due to 10 h irradiation can be attributed to the introduction of defects such as dislocations and vacancies similar to the variation of magnetization and coercivity in hysteresis graphs in **Figure 8**. Magnetization vs. temperature measurement for a wider range of temperatures between 4K and 350K, please refer to Figure S1 in supplemental material. Magnetization as a function of magnetic field isotherms between 250 K and 350 K of pre- and post-irradiated samples are also presented in the supplemental information (Figure S2).

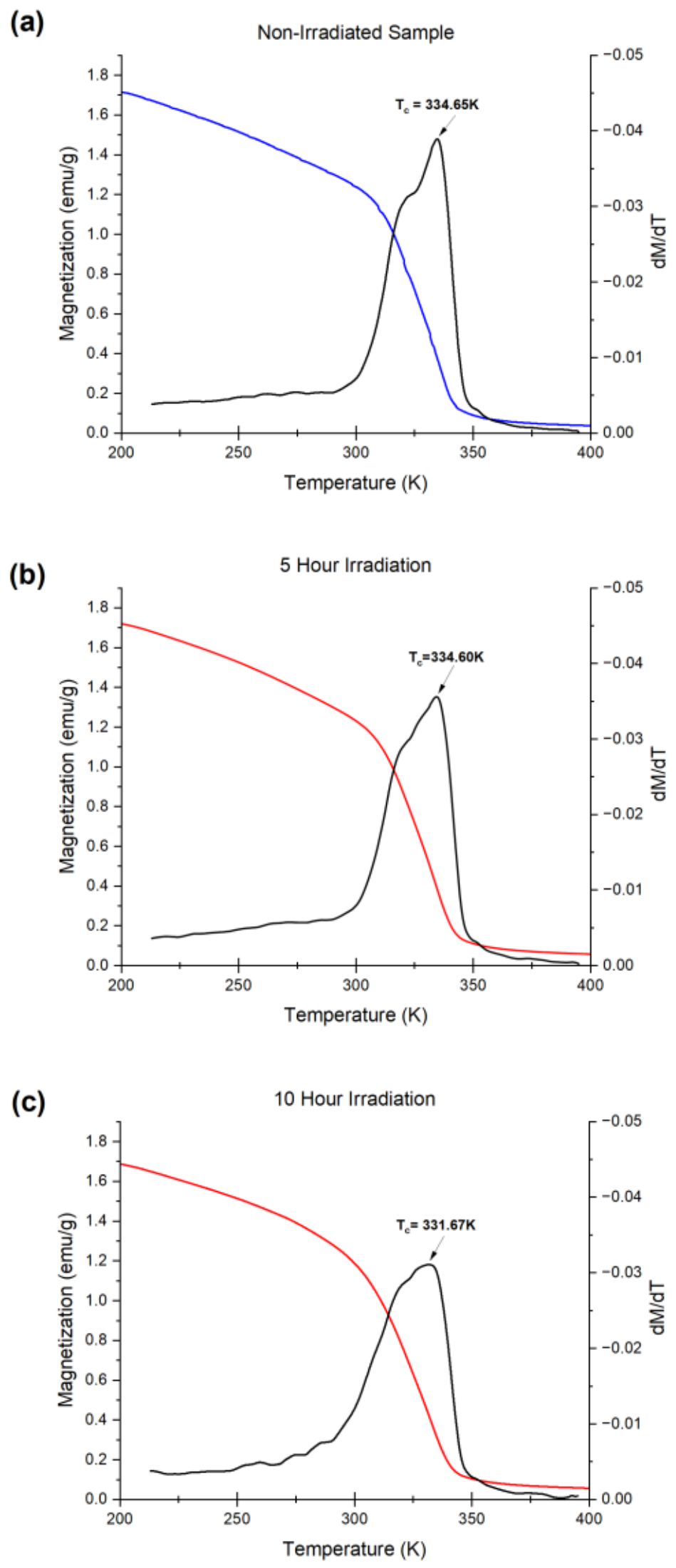

**Figure 9.** *Magnetization (M) as a function of temperature (T) at 100 Oe applied magnetic field when measured as (a) non-irradiated, and after X-ray irradiation for(b) 5h and (c) 10h of. The transition temperature was determined from dM/dT.*

The $Gd_5Si_4$ magnetic hyperthermia nanoparticles also exhibit a magnetocaloric effect near the magnetic phase transition temperature (≈ 336) K. Although the magnetocaloric effect will be small for the small AC magnetic field used in the hyperthermia treatment, it is still necessary to investigate the effect of radiation on the magnetocaloric effect in terms of isothermal entropy change. This material system exhibits other extreme property changes, such as colossal strain and giant magnetoresistance at the transition temperature, which may result in an increase in the number of defects [41], [42] [43]. Hence, it is crucial to investigate the temperature-dependent magnetic entropy change (ΔS). **Figure 10**, was generated using a signal processing technique involving an FFT filter with a cutoff frequency of 0.024 to effectively smooth the curve. The evaluation was conducted at a magnetic field of 3T using Maxwell's thermodynamic equations, expressed as follows:

$$\Delta S_M = \int_0^H \frac{\delta M}{\delta T} dH \qquad (2)$$

The curves show isothermal entropy changes for a magnetic field change from 0 to 3T, revealing a pronounced peak near 300 K. This peak shift from the transition temperature of 336K for bulk sample agrees with the previous literature [20] for micro and nanoparticles of $Gd_5Si_4$. This can be explained by the broader transition temperature in nanoparticles compared to a sharp phase transition in bulk materials leading to a peak shift in broadening in entropy calculations.

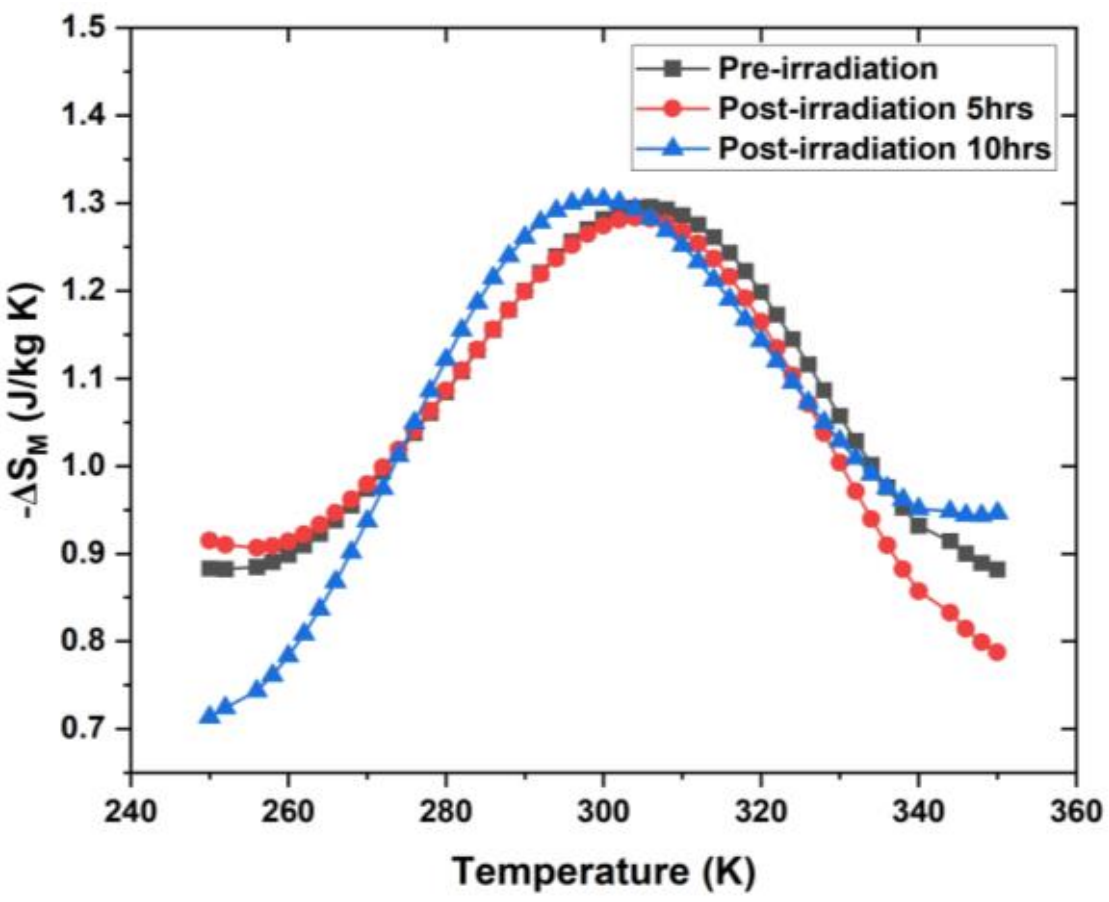

**Figure 10.** *Temperature-dependent magnetic entropy change ($\Delta S_M$) of pre- and post-irradiated samples for 5h and 10h combined graphs.*

The values of magnetic entropy change are significantly smaller compared to the bulk samples and occur at a wide temperature range, which also agrees with previous literature. The obtained maximum value is 1.30454 J/kg·K at a 3 T field change for pre- and post-irradiated samples and the position of the maximum value of around 304 K. The irradiation does not affect either the temperature at which the peak occurs or the peak itself.

### 3.6. Magnetic Hyperthermia Measurements

There was no significant difference in the SLP between the three samples, and the SLP was consistent with past measurements of polydispersed, non-irradiated $Gd_5Si_4$ particles [44], [45]. While this study did not investigate the self-regulating nature of the particles, past studies of polydispersed, non-irradiated $Gd_5Si_4$ particles, in which the field was applied for a longer period of time until a steady-state temperature was achieved, have shown a maximum temperature near the Curie temperature of the particles [44].

**Table 4.** *SLP of $Gd_5Si_4$ samples measured at 0.035 T and 227 kHz.*

| $Gd_5Si_4$ Sample | SLP (W/g) |
|---|---|
| **Non-irradiated** | 16.2 +/- 2.4 |
| **Irradiated 5h** | 17.3 +/- 2.6 |
| **Irradiated 10h** | 15.9 +/- 2.4 |

The SLP is proportional to the area enclosed within the hysteresis loop *M(H)* at the frequency used. Although some changes to the magnetization and coercivity are seen in the DC magnetization data after irradiation, these effects do not translate to significant differences in the SLP, within the uncertainty of the measurement. This may be due to differences in the magnetic behavior measured in the VSM (in the quasistatic, or DC measurement regime) compared to the magnetic behavior at 227 kHz used for the SLP measurements and in the frequency range used for MHT.

## 4. Conclusions

This study concludes that gadolinium silicide ($Gd_5Si_4$) nanoparticles exhibit stability in their magnetic and structural properties when exposed to high-dose (up to 72 kGy) and high-dose-rate (120 Gy/min) X-ray irradiation. Despite localized lattice distortions and dislocations observed via transmission electron microscopy, the nanoparticles retain their crystal structure, morphology, and elemental composition, as confirmed by X-ray diffraction, scanning electron microscopy, and energy-dispersive X-ray spectroscopy. Importantly, their self-regulating magnetic hyperthermia heating capability remains unaffected, with no significant changes in magnetization behavior or magnetocaloric properties, including isothermal entropy change values. These findings demonstrate the stability of $Gd_5Si_4$ nanoparticles under high-dose irradiation, supporting their potential application in combined therapeutic modalities such as magnetic hyperthermia and irradiation therapy. Furthermore, their resilience in high-energy, high-intensity radiation fields highlights their potential use in nuclear environments.

## Acknowledgments

Authors thank the funding from the Commonwealth Cyber Initiative Grant, NSF #2304513 and #2349694. The SLP measurements were supported by NASA under award no. 80NSSC20M0097, The Pennsylvania Space Grant Consortium (PSGC).